\documentclass{article}
\usepackage{arxiv}

\usepackage{amsmath}
\usepackage{amssymb}
\usepackage[version=3]{mhchem}
\usepackage{booktabs}
\usepackage{multirow}
\usepackage{graphicx}
\usepackage{float}
\usepackage{caption}
\usepackage{cite}
\usepackage{threeparttable}
\usepackage{amsfonts}
\usepackage{bm}
\usepackage[utf8]{inputenc}
\usepackage[T1]{fontenc}
\usepackage{hyperref}
\usepackage{url}
\usepackage{nicefrac}
\usepackage{microtype}
\usepackage{cleveref}
\usepackage{doi}

\AtBeginEnvironment{tablenotes}{\footnotesize}

\title{SpinMultiNet: Neural Network Potential Incorporating Spin Degrees of Freedom with Multi-Task Learning}
\author{
  Koki Ueno \\
	Panasonic Holdings Corporation\\
	Technology Division\\
	Osaka 571-8508 \\
	\texttt{ueno.koki@jp.panasonic.com} \\
	\And
  Satoru Ohuchi \\
	Panasonic Holdings Corporation\\
	Technology Division\\
	Osaka 571-8508 \\
	\And
  Kazuhide Ichikawa \\
	Panasonic Holdings Corporation\\
	Technology Division\\
	Osaka 571-8508 \\
	\And
  Kei Amii \\
	Panasonic Holdings Corporation\\
	Technology Division\\
	Osaka 571-8508 \\
	\And
  Kensuke Wakasugi \\
	Panasonic Holdings Corporation\\
	Technology Division\\
	Osaka 571-8508 \\
}

\date{\today}

\renewcommand{\headeright}{}
\renewcommand{\undertitle}{}

\begin{document}
\maketitle

\begin{abstract}
Neural Network Potentials (NNPs) have attracted significant attention as a method for accelerating density functional theory (DFT) calculations. However, conventional NNP models typically do not incorporate spin degrees of freedom, limiting their applicability to systems where spin states critically influence material properties, such as transition metal oxides. This study introduces SpinMultiNet, a NNP model that integrates spin degrees of freedom through multi-task learning. SpinMultiNet achieves accurate predictions without relying on correct spin values obtained from DFT or spin optimization calculations. Instead, it utilizes initial spin estimates as input and leverages multi-task learning to optimize the spin latent representation while maintaining both $E(3)$ and time-reversal equivariance. Validation on a dataset of transition metal oxides demonstrates the high predictive accuracy of SpinMultiNet. The model successfully reproduces the energy ordering of stable spin configurations originating from superexchange interactions and accurately captures the rhombohedral distortion of the rocksalt structure. These results pave the way for new possibilities in materials simulations that consider spin degrees of freedom, promising future applications in large-scale simulations of various material systems, including magnetic materials.
\end{abstract}

\section{Introduction}
First-principles calculations based on Density Functional Theory (DFT) have been widely utilized as a powerful tool for understanding electronic structures and material properties~\cite{saal2013_dft_sim1, jain2016_dft_sim2, hafner2008_dft_sim3}. Although DFT calculations can accurately predict energies and forces acting on atoms, they are often hindered by high computational costs. This limitation can become a significant bottleneck, particularly for large-scale systems or long-time simulations.
To address this issue, Neural Network Potentials (NNPs) have emerged as a promising alternative to accelerate DFT calculations~\cite{behler2007_bp, takamoto2022_matlantis, batzner20223_nequip, chanussot2021_oc20, yang2024_mattersim, chen2022_m3gnet, merchant2023_gnome}. NNPs learn the relationship between atomic configurations and energies from data obtained through DFT calculations, enabling significant reduction in computational cost while maintaining accuracy comparable to DFT calculations. In particular, NNPs based on the Graph Neural Network (GNN) are well-suited for constructing accurate potential models, as they can effectively capture the local atomic environments~\cite{batzner20223_nequip, schutt2017_schnet, gasteiger2021_gemnet, gasteiger2020_dimenet}.

However, most conventional NNPs do not account for spin degrees of freedom, limiting their application to material systems where spin states play a critical role in determining properties, such as transition metal oxides (TMOs). TMOs are known to exhibit diverse magnetic properties due to the presence of transition metal ions with partially filled d-orbitals, and incorporating spin degrees of freedom is crucial for understanding their properties~\cite{matar2003_tmo_calc}. For example, accurate prediction of the energy difference between ferromagnetic (FM) and antiferromagnetic (AFM) states requires proper representation of the potential energy surface depending on the spin configuration.
Recently, several NNP models incorporating spin degrees of freedom have been proposed~\cite{novikov2022_mMTP, eckhoff2021_sACSF, li2023_xdeeph}. These models can predict spin-dependent potential energy by using not only atomic coordinates but also spin values as inputs. However, to fully exploit the potential of these architectures, correct spin information is required as input. In practical situations, obtaining such correct spin values is often challenging, which hinders the utilization of these models.

To overcome this limitation, this study presents SpinMultiNet, a NNP model which utilizes initial spin estimates as input and accurately predicts the spin-dependent potential energy surface. Our model employs multi-task learning to predict energy, forces, and spin simultaneously. This allows the spin latent features to be optimized within the network, enabling highly accurate predictions, even if the input spin is an initial estimate provided by the user. Furthermore, our model is designed to satisfy not only $E(3)$ equivariance but also time-reversal equivariance, which ensures consistent and physically meaningful predictions. These equivariance contribute to improved data efficiency and enhanced generalization capabilities. The main contributions of this work are as follows:
\begin{enumerate}
  \item Development of a spin-dependent NNP model using initial spin estimates as input: We designed a spin-dependent NNP model applicable even when correct spin information is not available a priori.
  \item Demonstration of high prediction accuracy in TMOs: We applied SpinMultiNet to a dataset of TMOs and demonstrated its ability to accurately predict energy changes due to spin configurations. Specifically, we reproduced the energy ordering of stable spin configurations originating from superexchange interactions and confirmed that the optimized lattice constants of rocksalt TMOs agree well with experimental results.
  \item Verification of the importance of time-reversal equivariance: Ablation studies revealed that time-reversal equivariance is essential for accurate spin prediction. Additionally, we demonstrated that higher predictive accuracy can be achieved when precise spin values are provided as input.
\end{enumerate}

\section{Related Work}
In recent years, several NNP models that take spin degrees of freedom into account have been proposed. Magnetic moment tensor potential~\cite{novikov2022_mMTP} introduces spin degrees of freedom into the moment tensor potential~\cite{shapeev2016_MTP}, enabling the learning of spin-dependent potentials. Similarly, mHDNNP~\cite{eckhoff2021_sACSF} proposes a model that incorporates spin interactions into atom-centered symmetry functions. However, these methods are limited to collinear spins. 
On the other hand, SpinGNN~\cite{yu2024_spingnn} addresses noncollinear spins by extracting scalar features from two noncollinear spin vectors and using these features as input to the GNN. SpinGNN leverages the high expressive power of GNNs to construct accurate potential models.
These models have achieved success in several material systems, demonstrating the capabilities of spin-dependent NNPs.

A critical aspect of spin-dependent NNP models is ensuring time-reversal equivariance. Time-reversal equivariance describes how the state of a system changes under the time-reversal operation and is essential for the physically accurate handling of latent spin features. For example, under the time-reversal operation, spin flips its sign while energy remains invariant.
SpinGNN ensures the time-reversal invariance of the energy by restricting spin-derived features to time-reversal scalars only. On the other hand, xDeepH~\cite{li2023_xdeeph} achieves physical consistency and higher representational capacity by designing an architecture that is equivariant to the time-reversal operation.

However, most spin-dependent NNP models face the challenge of requiring correct spin values as input during prediction in order to maximize the performance of the architecture.
Correct spin values can be obtained through methods such as DFT calculations, iterative energy optimization using magnetic forces~\cite{yuan2024_magnet}, or by employing another machine learning model. However, executing these approaches for each NNP prediction is computationally expensive.

In contrast, CHGNet~\cite{deng2023_chgnet} outputs magnetic moments without requiring spin values as input. However, CHGNet is highly dependent on the spin states used in the training data and cannot predict energies or magnetic moments for spin states not included in the training data. Moreover, it outputs the same energy for structures with the same atomic configuration but different spin configurations, making it unsuitable for tasks such as predicting the energy difference between FM and AFM states.

SpinMultiNet accurately predicts energies and spin values even from initial spin estimates, without relying on correct spin values obtained from DFT or spin optimization calculations.
This is achieved by performing multi-task learning that simultaneously predicts energy and spin while optimizing spin features in a time-reversal equivariant manner.
This approach enables efficient calculation of energies for various spin configurations, addressing the computational cost challenges of conventional methods.
It should be mentioned that our approach is very similar to the recently proposed SpinGNN++\cite{yu2022_spingnnpp}. However, this study specifically focuses on investigating the influence of spin prediction from initial estimates as a supervisory signal on the overall performance of the model.

\section{Methods}
\subsection{Equivariance}
\subsubsection{$E(3)$ Equivariance}
Equivariance refers to the property where the output changes correspondingly when a specific transformation is applied to the input data. For instance, if the input data is rotated, the output of an equivariant function will also rotate accordingly. This property plays a crucial role in processing physical systems and geometric data.
Generally, a function $\mathcal{L}: \mathcal{X} \rightarrow \mathcal{Y}$ ($\mathcal{X}$, $\mathcal{Y}$ are vector spaces) is equivariant if the representation $D$ of the group $G$ satisfies the following:
\begin{equation}
  \mathcal{L} \circ D^{\mathcal{X}}(g) = D^{\mathcal{Y}}(g) \circ \mathcal{L}
\end{equation}
Here, $D^{\mathcal{X}}(g)$ is the representation of the vector space $\mathcal{X}$ for element $g$ of the group $G$.
$SO(3)$ equivariance refers to the property of being equivariant to rotation operations in three-dimensional space.
The irreducible representations of $SO(3)$ are known as Wigner D-matrices~\cite{gilmore2008_lie}, which are matrices of dimension $2l + 1$ for rotation order $l$. By incorporating $SO(3)$ equivariance into each layer of the model, an overall $SO(3)$ equivariant model can be constructed. This is achieved by combining two steerable vector features using the Clebsch-Gordan tensor product~\cite{thomas2018_TFN, brandstetter2021_steerable}.
\begin{equation}\label{eq:tensor_prod}
  (\mathbf{u} \otimes \mathbf{v})^{(l)}_m = \sum_{m_1=-l_1}^{l_1} \sum_{m_2=-l_2}^{l_2} C^{(l,m)}_{(l_1,m_1)(l_2,m_2)} u^{(l_1)}_{m_1} v^{(l_2)}_{m_2}
\end{equation}
Here, $\mathbf{u}$ and $\mathbf{v}$ are steerable vector features of rotation orders $l_1$ and $l_2$, respectively, $m$ is the representation index ($m \in [-l, l]$), and $C^{(l,m)}_{(l_1,m_1)(l_2,m_2)}$ is the Clebsch-Gordan coefficient.
A steerable vector feature is a $2l + 1$ dimensional vector that takes the form of an irreducible representation of the $SO(3)$ group and can be rotated by applying the Wigner D-matrix~\cite{brandstetter2021_steerable}.
This tensor product has a non-zero value only when $l$ satisfies $|l_1 - l_2| \leq l \leq |l_1 + l_2|$, and the output is also an irreducible representation.
Furthermore, by calculating feature vectors using the interatomic vector $\vec{r}_{ij}$ and restricting the tensor product calculation to cases where the parity $p$ ($-1$ for odd and $1$ for even) satisfies the condition $p_l = p_{l_1}p_{l_2}$, it is possible to construct an $E(3)$ equivariant model that incorporates translation and inversion operations~\cite{geiger2022_e3nn}.
$E(3)$ equivariant models can flexibly represent interactions between scalars, vectors, and higher-order tensors, leading to high expressive power in processing data in three-dimensional space and improved data efficiency.

\subsubsection{Time-Reversal Equivariance}
Since spin changes its sign under the time-reversal operations, it is important to incorporate this equivariance into the NNP model.
According to xDeepH~\cite{li2023_xdeeph}, introducing time-reversal equivariance $\{I, \mathcal{T}\}$ into the $E(3)$ equivariant model can be achieved by decomposing the tensor product of two spin vectors as $\frac{1}{2} \otimes \frac{1}{2} = 0 \oplus 1$.
Time-reversal equivariance is then achieved by ensuring that the imaginary part of $l=0$ and the real part of $l=1$ change their sign under the time-reversal operation, while other components remain unchanged.
In practice, this can be incorporated into Equation \eqref{eq:tensor_prod} by introducing a time-reversal parity $t$. Each vector feature is labeled with four parameters: $l$, $m$, $p$, and $t$. For example, $\vec{r}_{ij}$ is labeled as ($l=1$, $p=-1$, $t=1$), and $\vec{m}_{i}$ as ($l=1$, $p=1$, $t=-1$).
$t$ is treated similarly to $p$, and the tensor product is calculated only when the condition $t_l = t_{l_1}t_{l_2}$ is satisfied.
Additionally, scalar spin features with ($l=0$, $p=1$, $t=1$), i.e., $E(3) \times \{I, \mathcal{T}\}$ invariant features, can also be incorporated.
In this study, we added the magnitude of the magnetic moment to the initial node features and the inner product of magnetic moments between neighboring atoms to the edge features.
By incorporating time-reversal equivariance in this manner, we expect the model to represent physically correct spin behavior, leading to improved prediction accuracy.

\subsection{Model Architecture}
\begin{figure}
\centering
\includegraphics[width=\linewidth]{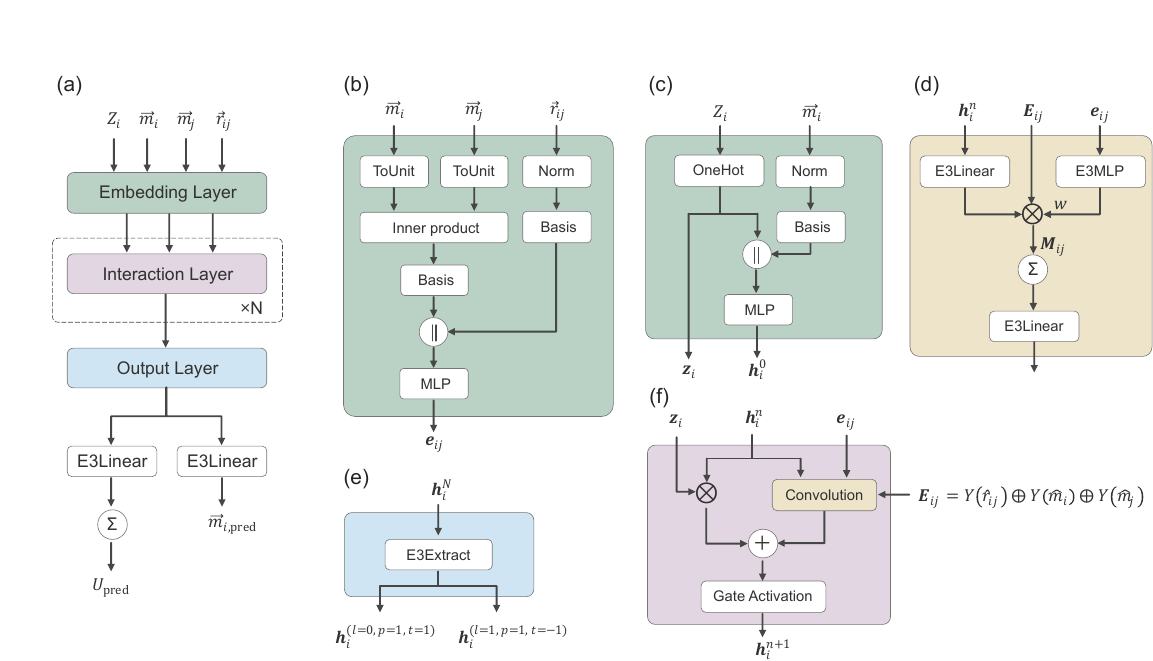}
\caption{\label{fig:arch} Overall model architecture.
(a) Input and output of the model. The atomic number $Z$, the initial magnetic moment estimate $\vec{m}$, and interatomic vectors $\vec{r}_{ij}$ are input and processed through the Embedding Layer, $N$ Interaction Layers, and the Output Layer to output the energy $U_{\text{pred}}$ and magnetic moments $\vec{m}_{i, \text{pred}}$.
(b) Embedding Layer (edge). Edge features are created using the magnetic moments of two atoms and the interatomic vector. \texttt{ToUnit} represents the operation of converting a vector to a unit vector, and $||$ represents the concatenation of tensors.
(c) Embedding Layer (node). Element embedding vectors and initial node features are created using the atomic number and the magnetic moment.
(d) Convolution. Messages are created and aggregated using node and edge features.
(e) Output Layer. Steerable features for energy and magnetic moment predictions are extracted from the latent features.
(f) Interaction Layer. Node features are updated. $Y$ represents the expansion using spherical harmonics, and $\otimes$ represents the tensor product.
}
\end{figure}

SpinMultiNet is built upon a GNN. Figure~\ref{fig:arch} illustrates the overall architecture of the model. Note that in this paper, any process labeled with E3 represents an $E(3) \times \{I, \mathcal{T}\}$ equivariant process. These processes were implemented using the \texttt{e3nn}\cite{geiger2022_e3nn} and \texttt{xdeeph}\cite{li2023_xdeeph} packages.
For each atom, steerable features are generated using the atomic number $Z_i$, interatomic vectors $\vec{r}_{ij}$, and initial magnetic moment estimate $\vec{m}_i$ as input. These features are then fed into an $E(3) \times \{I, \mathcal{T}\}$ equivariant GNN. The steerable features are iteratively updated by the Interaction Layers, after which specific irreducible representations are extracted and used for predicting the energy and magnetic moments.
While forces acting on atoms can be directly predicted from the $l=1$ features, in this study, they were calculated from the gradient of energy with respect to atomic positions.
The Interaction Layer is designed to be $E(3) \times \{I, \mathcal{T}\}$ equivariant, ensuring that the node features of each atom are updated while maintaining equivariance. By stacking multiple Interaction Layers, SpinMultiNet can capture longer-range atomic and spin interactions.

\subsubsection{Embedding Layer}
Unlike conventional models that do not consider spin degrees of freedom, our model incorporates spin information into the atom embedding. The initial node features $\mathbf{h}_i^0$ and edge features $\mathbf{e}_{ij}$ are defined as follows:
\begin{equation}
  \mathbf{h}_i^0 = \text{MLP}(\text{OneHot}(Z_i)\ ||\ \text{GaussianBasis}(|\vec{m}_i|))
\end{equation}
\begin{equation}
  \mathbf{e}_{ij} = \text{MLP}(\text{BesselBasis}(|\vec{r}_{ij}|)\ ||\ \text{GaussianBasis}(\hat{m}_i \cdot \hat{m}_j))
\end{equation}
Here, $\hat{m}_i$ represents the unit vector of the initial magnetic moment estimate. A cutoff function is applied to the Bessel functions to ensure smoothness before and after the cutoff~\cite{gasteiger2020_dimenet}.
Spin information is incorporated into the initial node and edge features by concatenating scalar features that are invariant under the time-reversal operation (the magnitude of the magnetic moment and the inner product of two magnetic moments, respectively).

\subsubsection{Interaction Layer}
The $E(3) \times \{I, \mathcal{T}\}$ equivariant operation used as the convolution layer in SpinMultiNet is defined as follows:
\begin{align}
  \begin{split}
    (M_{ij, c})^{(l)}_m
    &= \left( \mathbf{h}^{(l_1)}_{j,c} \otimes^w \mathbf{E}^{(l_2)}(\vec{r}_{ij}, \vec{m}_{i}, \vec{m}_{j}) \right)^{(l)}_m\\
    &= w \sum_{m_1=-l_1}^{l_1} \sum_{m_2=-l_2}^{l_2} C^{(l,m)}_{(l_1,m_1)(l_2,m_2)} h^{(l_1)}_{j,c, m_1} E^{(l_2)}_{m_2}(\vec{r}_{ij}, \vec{m}_{i}, \vec{m}_{j})
  \end{split}
\end{align}
\begin{equation}
  \mathbf{E}(\hat{r}_{ij}, \hat{m}_{i}, \hat{m}_{j}) = Y(\hat{r}_{ij}) \oplus Y(\hat{m}_i) \oplus Y(\hat{m}_j)
\end{equation}
\begin{equation}
  \mathbf{w} = \text{E3MLP}(\mathbf{e}_{ij})
\end{equation}

Here, $(M_{ij, c})^{(l)}_m$ is the $l$, $m$ element of channel $c$ in the message function, $\mathbf{h}$ is the latent feature of the node, and $Y$ is the spherical harmonics function. For brevity, parities $p$ and $t$ are omitted.
$\mathbf{w}$ is a weight vector that has a value for each path of the tensor product and is calculated from the edge features $\mathbf{e}_{ij}$. To satisfy equivariance, the same weight must be applied for the same $l$ regardless of the value of $m$.
Each feature has a parity $t$ with respect to the time-reversal operation, and directional information of the magnetic moment is incorporated through $\mathbf{E}$. An $E(3) \times \{I, \mathcal{T}\}$ equivariant tensor product is calculated between the node features and $\mathbf{E}$, and weighted by the value calculated from the edge features. Through this process, the model can capture spatial and spin interactions between atoms. When considering only collinear spin, there is no need to expand the magnetic moment in spherical harmonics; it can simply be input as a time-reversal odd scalar.
The calculated message is aggregated to the central node through the message function, and a non-linear activation function is applied. In this study, a gate-type activation function with time-reversal equivariance~\cite{geiger2022_e3nn, li2023_xdeeph} was used. Figure~\ref{fig:equivariance} shows the visualization of the changes in latent features with respect to input structure rotation and spin inversion.

\subsubsection{Output Layer}
After passing through multiple Interaction Layers, the node features retain sufficient information regarding the structure and spin configurations. From these node features, the energy and magnetic moments are calculated.
The energy is calculated by a linear combination of the components of each node feature that satisfy the following conditions: rotation order $l=0$, parity $p=1$, and time-reversal parity $t=1$. These components are essentially the $E(3) \times \{I, \mathcal{T}\}$ invariant scalar components.
\begin{equation}
  U_{\text{pred}} = \sum_i \sum_c w_c h_{i, c}^{(l=0, p=1, t=1)}
\end{equation}
Here, $w_c$ represents the learnable weight parameters, and $h_{i, c}^{(l=0, p=1, t=1)}$ represents the $E(3) \times \{I, \mathcal{T}\}$ invariant scalar component of the node features of atom $i$.
On the other hand, the magnetic moment is calculated by a linear combination of the components of each node feature that satisfy $l=1$, $p=1$, and $t=-1$. These components represent the latent spin representation reflecting the structural information.
\begin{equation}
  \vec{m}_{i,\ \text{pred}} =\sum_c w_c \mathbf{h}_{i, c}^{(l=1, p=1, t=-1)}
\end{equation}
Here, $\vec{m}_{i,\ \text{pred}}$ represents the predicted magnetic moment of atom $i$, and $\mathbf{h}_{i, c}^{(l=1, p=1, t=-1)}$ represents the latent spin representation of the node features of atom $i$.
For collinear spins, the components with $l=0$, $p=1$, and $t=-1$ are used.
We performed multi-task learning by introducing an auxiliary task of predicting the magnetic moments obtained from DFT calculations, in addition to predicting the energy. This allows the model to learn correct spin information internally and improve the energy prediction accuracy, even if the input magnetic moments are initial estimates.
The loss function for multi-task learning is defined as a weighted sum of the losses for energy, force, and magnetic moment predictions.
\begin{equation}\label{eq:loss_func}
  \mathcal{L} = \mathcal{L}_{\text{energy}} + \lambda_f \mathcal{L}_{\text{forces}} + \lambda_m \mathcal{L}_{\text{mag}}
\end{equation}
Here, $\mathcal{L}_{\text{energy}}$, $\mathcal{L}_{\text{forces}}$, and $\mathcal{L}_{\text{mag}}$ represent the loss functions for energy, force, and magnetic moment predictions, respectively, and $\lambda_f$ and $\lambda_m$ represent the weight coefficients for each loss. In this study, $\mathcal{L}_{\text{mag}}$ was applied only to magnetic elements.
Through this multi-task learning, the model can predict the energy and magnetic moments corresponding to various spin states, such as ferromagnetic and antiferromagnetic states, and high-spin and low-spin states, by modifying the direction and magnitude of the initial estimates of the input magnetic moments. This mimics the calculation process of first-principles calculation software such as Vienna Ab-initio Simulation Package (VASP)~\cite{blochl1994_paw}, indicating that the input and output results of VASP can be directly used as training data.

\subsection{Dataset}
The datasets were created using DFT calculations performed with VASP. The detailed settings for the DFT calculations are provided in Appendix~\ref{sec:DFT}.
First, we created a dataset (Mn-Co-Ni dataset) focusing on rocksalt-type TMOs with the space group $\mathrm{Fm\overline{3}m}$.
Specifically, using Mn, Co, Ni, or their combinations as transition metal atoms, we performed structural optimizations starting from FM and various AFM configurations to obtain stable crystal structures. Following this, we applied four types of deformation operations to build a dataset that includes a diverse range of atomic configurations:
\begin{itemize}
  \item Random displacement: Each atomic coordinate was randomly displaced by a small amount.
  \item Shear strain: Shear strain was applied while maintaining the fractional coordinates of the atoms.
  \item Tensile strain: Tensile strain was applied while maintaining the fractional coordinates of the atoms.
  \item Cell volume change: The volume of the crystal lattice was changed while maintaining the fractional coordinates of the atoms.
\end{itemize}
For each structure with these deformation operations applied, we performed single-point calculations using VASP to calculate the energy, forces, and magnetic moments. The magnetic moments were directly used from the VASP output. Each data point also includes the initial magnetic moments (\texttt{MAGMOM}) obtained from the VASP input file, which are used as inputs to the NNP model.
Finally, we constructed the Mn-Co-Ni dataset, consisting of a total of 29,989 data points.
Additionally, we created a dataset focused on the \ce{CoO} crystal structure (space group $\mathrm{Fm\overline{3}m}$), referred to as the Co-pair dataset. For 1,000 structures generated by applying random displacements, both FM (ferromagnetic) and AFM (antiferromagnetic) configurations were generated, resulting in 1,000 pairs of structures. In each structure pair, only the spin configurations differ.
Similarly, for these structure pairs, single-point calculations were performed to obtain the training data.
The Co-pair dataset is used to learn the energy difference between different spin configurations for the same atomic configuration.
The Mn-Co-Ni dataset was randomly split into training, validation, and test sets with a ratio of 80\%, 10\%, and 10\%, respectively. The Co-pair dataset was similarly split, ensuring that each structure pair belongs to only one of the splits.
Using these datasets, we trained SpinMultiNet to minimize the loss function defined in Equation~\eqref{eq:loss_func}. The detailed settings for the training are provided in Appendix~\ref{sec:training}.

\section{Results}
\subsection{Model Performance}
First, we present the mean absolute errors (MAEs) for the Mn-Co-Ni dataset in Table~\ref{tab:MnCoNi_res}.
\begin{table}[t]
\centering
\caption{MAEs for the Mn-Co-Ni dataset.}
\label{tab:MnCoNi_res}
\begin{tabular}{lccc}
\toprule
Model & Energy (meV/atom) & Forces (meV/\AA) & Magnetic moments ($\mu_B$) \\
\midrule
w/o spin (NequIP) & 8.322 & 9.956 & -- \\
SpinMultiNet (single-task) & 2.262 & 8.348 & -- \\
SpinMultiNet (multi-task) & 2.229 & 8.172 & 0.0076 \\
\bottomrule
\end{tabular}
\end{table}
\begin{figure}[t]
\centering
\includegraphics[width=\linewidth]{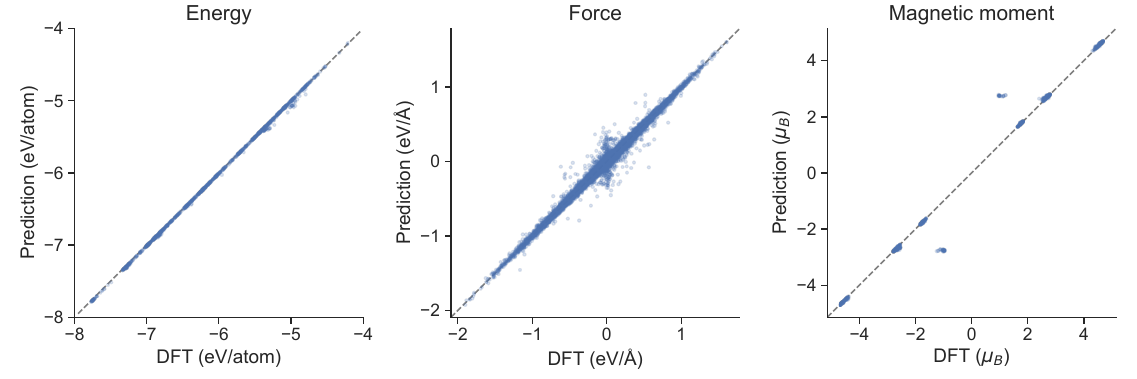}
\caption{\label{fig:MnCoNi_scatter} Scatter plots of predicted values versus DFT calculated values for each property in the Mn-Co-Ni dataset. (a) Energy, (b) Forces, (c) Magnetic moments.}
\end{figure}
For comparison, we also show the results of an NNP model without spin input (NequIP~\cite{batzner20223_nequip}), with a comparable number of training parameters.
SpinMultiNet showed an improvement in prediction accuracy of 73.2\% for energy and 17.9\% for forces compared to the model without spin input.
Furthermore, when performing multi-task learning that includes spin output, the prediction accuracy for both energy and forces improved even further compared to single-task learning, despite the optimization cost being allocated to the magnetic moment as well.
This finding suggests that predicting magnetic moments refines the latent representation of input spins, aligning it more closely with the correct spin information obtained from DFT calculations, thereby improving energy prediction accuracy.
Moreover, multi-task learning enables the prediction of magnetic moments during inference. As shown in Figure~\ref{fig:MnCoNi_scatter}(c), the model provides accurate predictions of magnetic moments. However, large prediction errors were observed for some atoms. These errors stem from the misprediction of low-spin \ce{Co^{2+}} species as high-spin states, likely due to the limited number of low-spin \ce{Co^{2+}} species in the dataset. Nevertheless, these instances account for only 0.3\% of the total magnetic atoms, indicating that the model accurately predicts magnetic moments for the majority of atoms.

In the Mn-Co-Ni dataset, all structures exhibit slight variations in atomic configurations. Therefore, even a model lacking spin input may be capable of inferring the spin state to some degree based on these structural differences, which could, in turn, reduce the energy prediction error.
To more clearly verify the effect of spin input, we conducted additional experiments using the Co-pair dataset.
The Co-pair dataset contains energy data for both FM and AFM configurations of the same atomic configurations, enabling a clearer demonstration of the importance of the spin input.
Table~\ref{tab:Co_res} shows the MAEs for the Co-pair dataset. The model without spin input exhibits a significant energy prediction error of 26.6~meV/atom and predicts an intermediate energy between the FM and AFM states for all data points. This is a reasonable result, as the model without spin input cannot distinguish between different spin configurations. Conversely, SpinMultiNet demonstrates a very small energy prediction error of 0.403~meV/atom, confirming its ability to clearly distinguish between FM and AFM states.
\begin{table}[t]
\centering
\caption{MAEs for the Co-pair dataset.}
\label{tab:Co_res}
\begin{tabular}{lccc}
\toprule
Model & Energy (meV/atom) & Forces (meV/\AA) & Magnetic moments ($\mu_B$) \\
\midrule
w/o spin (NequIP) & 26.632 & 30.156 & -- \\
SpinMultiNet & 0.403 & 1.846 & 0.0018 \\
\bottomrule
\end{tabular}
\end{table}
These results demonstrate that by appropriately considering spin degrees of freedom, SpinMultiNet can predict energy, forces, and magnetic moments with higher accuracy compared to conventional NNP models without spin input.

\subsection{Identification of Stable Spin Configurations}
\begin{figure}[t]
\centering
\includegraphics[width=\linewidth]{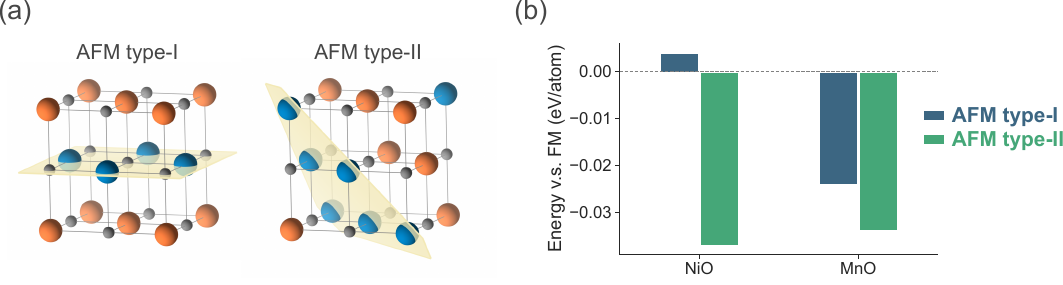}
\caption{\label{fig:afm} (a) Two types of antiferromagnetic structures. In AFM type-I, spins are aligned parallel within the (001) plane, while in AFM type-II, spins are aligned parallel within the (111) plane. (b) Energy comparison after structural optimization. The energy difference relative to the FM configuration is plotted.}
\end{figure}

SpinMultiNet can predict the energy for any given spin configuration, enabling the identification of the most stable spin configuration in a magnetic structure. In this section, we performed structural optimizations for NiO and MnO with $\mathrm{Fm\overline{3}m}$ rocksalt structures, using FM and two types of AFM configurations (AFM type-I and AFM type-II shown in Figure~\ref{fig:afm}(a), visualized using VESTA~\cite{momma2011_vesta}) as initial structures to predict the most stable spin configuration. The structural optimization were performed without symmetry constraints using \texttt{ASE}~\cite{ase_gitlab}.
Figure~\ref{fig:afm}(b) shows the energy after structural optimization for each spin configuration. For both NiO and MnO, the AFM type-II configuration was identified as the most stable spin configuration. This agrees with the experimentally observed antiferromagnetic ground state~\cite{slack1960_afm_typeii, cheetham1983_afm_typeii_NiMnO}.
In these TMOs, it is known that due to superexchange interactions, the AFM type-II configuration, where spins align parallel within the (111) plane, is more stable than the AFM type-I configuration~\cite{anderson1950_superexchange1, anderson1963_superexchange2}. SpinMultiNet accurately reproduces this energy ordering originating from superexchange interactions.
Table~\ref{tab:lattice_consts} shows the lattice constants after structural optimization for each spin configuration. In the FM configuration, symmetry was preserved after structural optimization, whereas in the AFM type-I configuration, a distortion along the $c$-axis was observed. Notably, a rhombohedral distortion was induced in the AFM type-II configuration, altering the space group to $\mathrm{R\overline{3}m}$. The optimized rhombohedral angle $\alpha$ (= $\beta$ = $\gamma$) was $90.10^\circ$ and $90.64^\circ$ for NiO and MnO, respectively. These values are in excellent agreement with the experimentally reported rhombohedral angles ($90.08^\circ$ for NiO and $90.60^\circ$ for MnO)~\cite{cheetham1983_afm_typeii_NiMnO}.
\begin{table}[t]
\centering
\caption{Lattice constants after structural optimization for each structure and spin configuration. Experimental values are taken from Ref.~\cite{cheetham1983_afm_typeii_NiMnO}.}
\label{tab:lattice_consts}
\begin{tabular}{llccccccc}
\toprule
& & & \multicolumn{5}{c}{Lattice constant} \\
\cmidrule(lr){4-8}
Material & Ordering & Energy (eV/atom) & $a$ & $b$ & $c$ & $\alpha$ (= $\beta$ = $\gamma$) & $\alpha$ (experiment) \\
\midrule
\multirow{3}{*}{NiO} & FM & -4.809 & 8.492 & 8.492 & 8.492 & 90.00 & - \\
& AFM type-I & -4.805 & 8.494 & 8.494 & 8.507 & 90.00 & - \\
& AFM type-II & \textbf{-4.846} & 8.455 & 8.455 & 8.455 & 90.10 & 90.08 \\
\midrule
\multirow{3}{*}{MnO} & FM & -7.742 & 9.042 & 9.042 & 9.042 & 90.00 & - \\
& AFM type-I & -7.766 & 9.024 & 9.024 & 8.988 & 90.00 & - \\
& AFM type-II & \textbf{-7.776} & 8.993 & 8.993 & 8.993 & 90.64 & 90.60 \\
\bottomrule
\end{tabular}
\end{table}
These results demonstrate that SpinMultiNet effectively learns the complex, spin-dependent energy landscape and can accurately predict both the stable spin configuration and the associated structural parameters.
This suggests that SpinMultiNet can be a powerful tool for exploring stable spin configurations in complex systems where DFT calculations are computationally expensive and challenging.

\subsection{Ablation Study}
To further understand the behavior of SpinMultiNet, we performed an ablation study on its architecture and input features. Table~\ref{tab:ablation} shows the results of the ablation study using the Mn-Co-Ni dataset.
\begin{table}[t]
\centering
\caption{Ablation study. MAEs for the Mn-Co-Ni dataset.}
\label{tab:ablation}
\begin{tabular}{lccc}
\toprule
Model & Energy (meV/atom) & Forces (meV/\AA) & Magnetic moments ($\mu_B$) \\
\midrule
Time-reversal invariant & 2.321 & 8.417 & 0.5840 \\
Accurate input spin & 1.257 & 7.574 & -- \\
\midrule
SpinMultiNet & 2.229 & 8.172 & 0.0076 \\
\bottomrule
\end{tabular}
\end{table}
First, to examine the effect of time-reversal equivariance in SpinMultiNet, we trained a version with the spin-related components removed from $\mathbf{E}(\hat{r}_{ij}, \hat{m}_{i}, \hat{m}_{j})$, making it time-reversal invariant.
The time-reversal invariant model showed a slight increase in MAE for energy and forces by 4.13\% and 3.0\%, respectively, compared to SpinMultiNet. However, the MAE for the magnetic moment increased significantly, from 0.0076~$\mu_B$ to 0.5840~$\mu_B$.
This is because the time-reversal invariant model cannot recognize the inversion of the input spin and incorrectly predicts the sign of the output spin. In contrast, SpinMultiNet (time-reversal equivariant) can correctly change the sign of the output spin and internal features in response to the inversion of the input spin.

Next, to investigate the performance when using correct input spins, we trained a model using the magnetic moments obtained from DFT calculations as input spins. In this case, since the input and output spins are identical, only the energy and forces were used as training targets.
This model achieved a reduction in MAE of 43.6\% for energy and 7.32\% for forces compared to the model using initial magnetic moment estimates as inputs.
This suggests that using more precise values for the input spins can further enhance the model performance.
Remarkably, even when using a single initial estimate for the magnetic moment of each element (in this study, 3.0 for Mn and 2.5 for Ni), SpinMultiNet demonstrates high performance. The results are comparable to those obtained using the correct magnetic moments, with the difference in the MAE of energy prediction being within 1~meV/atom.
This suggests that SpinMultiNet can accurately predict energies as long as the spin direction is correctly specified, indicating that determining the initial estimates is relatively straightforward.
The results of this ablation study highlight the importance of time-reversal equivariance and spin input values, supporting the validity of SpinMultiNet design.

\section{Conclusion}
In this study, we developed SpinMultiNet, a multitasking NNP model which explicitly incorporates spin degrees of freedom. This model can simultaneously predict accurate energies and spin values using initial spin estimates as input, without relying on correct spin values obtained from DFT calculations.
This was achieved by employing multi-task learning to simultaneously predict energy and spin, optimizing the latent representation of spin in the process.
SpinMultiNet accurately captures the spin-dependent energy landscape and can reproduce important physical phenomena such as superexchange interactions. This paves the way for large-scale simulations of various material systems, including magnetic materials, which were challenging for conventional NNP models.

Future challenges include the following two points:
\begin{itemize}
\item Validation using larger datasets: While we validated the effectiveness of the model using a relatively small dataset in this study, evaluation using a large and diverse dataset is necessary to verify its applicability to a wider range of material systems. Considering spin degrees of freedom increases the complexity of the energy landscape, requiring more training data than conventional NNP models.
\item Improvement of the mapping between initial estimates and converged values: For magnetic moments, the initial estimates and the converged values obtained from DFT calculations have a many-to-one relationship. This means that slightly different (or sometimes significantly different) initial magnetic moment estimates can correspond to the same converged value, complicating the model training. To address this issue, appropriate constraints need to be introduced into the model to learn a proper mapping between initial estimates and converged values.
\end{itemize}
By addressing these challenges, we expect to develop even more accurate and versatile spin-dependent NNP models.

\bibliographystyle{unsrt}
\bibliography{reference}

\newpage
\appendix
\numberwithin{equation}{section}
\numberwithin{figure}{section}
\numberwithin{table}{section}
\section*{Appendix}

\section{Visualization of Latent Features}\label{sec:visualization}
\begin{figure}[h]
\centering
\includegraphics[width=\linewidth]{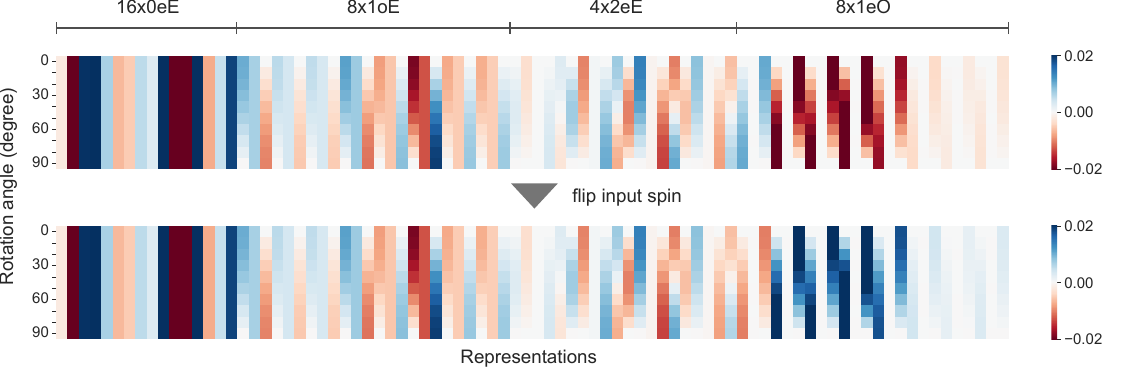}
\caption{\label{fig:equivariance} Visualization of the latent features of the Ni atom in a Ni-O two-atom system. The upper part represents the case where Ni has an up-spin, and the lower part represents the down-spin case. The vertical axis indicates the changes in features when the input structure is rotated.}
\end{figure}

Since SpinMultiNet consists of $E(3) \times \{I, \mathcal{T}\}$ equivariant interaction layers, its latent features also possess this equivariance.
To illustrate this behavior, Figure~\ref{fig:equivariance} shows the changes in the latent features of a Ni atom within a Ni-O two-atom system when the structure is rotated and the spin is flipped.
Here, the latent features have 16$\times$0eE + 8$\times$1oE + 4$\times$2eE + 8$\times$1eO representations. For example, 8$\times$1oE represents 8 channels of vector features with rotation order $l=1$, parity $p=-1$ (odd), and time-reversal parity $t=1$ (even).
The upper part of Figure~\ref{fig:equivariance} shows that while the features in the $l>0$ region rotate with the input structure, those in the $l=0$ region remain unchanged. This demonstrates that SpinMultiNet satisfies $E(3)$ equivariance.
Furthermore, the lower part of Figure~\ref{fig:equivariance} shows that when the input spin is flipped, the features in the $t=-1$ region (time-reversal odd features) are inverted, while those in the $t=1$ region (time-reversal even features) remain unchanged. This demonstrates that SpinMultiNet satisfies time-reversal equivariance.
Thus, the internal features of SpinMultiNet appropriately transform in response to changes in the input, enabling data-efficient learning.

\section{DFT Calculations}\label{sec:DFT}
Spin-polarized DFT calculations were performed using VASP. The Perdew-Burke-Ernzerhof (PBE) functional~\cite{perdew1996_pbe} was used as the exchange-correlation functional, and the calculations were based on the GGA+U method with Hubbard U correction. The $U-J$ parameters for Co, Ni, and Mn were set to 3.32~eV, 6.2~eV, and 3.9~eV, respectively.
The plane-wave cutoff energy was set to 520~eV. The $k$-point mesh was automatically generated using the \texttt{Kpoints.automatic\_density\_by\_vol} method implemented in the \texttt{pymatgen}~\cite{ong2013_pymatgen} package under the condition of $\text{kppvol} = 100$.
Single-point calculations were performed for each structure to calculate the energy, forces, and magnetic moments. The input magnetic moments were set to 1.0, 2.5, 3.0, and 0.0~$\mu_B$ for Co, Ni, Mn, and O, respectively, assuming collinear spins. These values were also used as initial estimates of magnetic moments for input into the SpinMultiNet.
It is important to note that these initial estimates are different from the final magnetic moments obtained through DFT calculations.
This accounts for the difficulty in obtaining correct magnetic moments in advance under realistic simulation scenarios.

\section{Training Details}\label{sec:training}
SpinMultiNet, with four Interaction Layers, was trained to minimize the loss function defined in Equation~\eqref{eq:loss_func}. The MAE was used as the loss function, and the loss weights for forces and magnetic moments were set to $\lambda_f = 1.0$ and $\lambda_m = 0.1$, respectively. The energy loss was calculated after converting to per-atom energy.
The Adam optimizer was used with an initial learning rate of 0.01, a batch size of 32, and 400 epochs. The learning rate was decayed to $1 \times 10^{-4}$ using a cosine annealing scheduler.
For comparison, we trained the NequIP model \cite{batzner20223_nequip}, which does not account for spin degrees of freedom, using the same settings. The number of model parameters was adjusted to be approximately the same as SpinMultiNet (about 4M), and the maximum rotation order was limited to $l=2$.
The training of these models was performed using NVIDIA V100 GPUs.

\end{document}